\documentclass[letterpaper,12pt]{JHEP3}
\usepackage{amsmath,amssymb}
\usepackage{epsfig}
\usepackage{cite}
\usepackage[english]{babel}

\newcommand{\dd}{{\rm d}}

\newcommand{\eq}{\begin{equation}}
\newcommand{\eeq}{\end{equation}}
\newcommand{\eqn}{\begin{eqnarray}}
\newcommand{\feqn}{\end{eqnarray}}
\newcommand{\arr}{\begin{eqnarray*}}
\newcommand{\farr}{\end{eqnarray*}}

\newcommand{\F}{{\cal F}}

\newcommand{\p}{\partial}
\newcommand{\w}{\wedge}

\newcommand{\D}{{\mathcal{D}}}
\newcommand{\FF}{{\mathcal{F}}}
\newcommand{\NN}{{\mathcal{N}}}

\newcommand{\R}{{\mathbb{R}}}

\newcommand{\V}{{\mathcal{V}}}

\def\al{\alpha}
\def\be{\beta}

\def\ep{\epsilon}

\def\la{\lambda}

\def\si{\sigma}

\def\Om{\Omega}

\title{Supersymmetric gyratons in five dimensions}
\author{Marco M. Caldarelli, Dietmar Klemm and Emanuele Zorzan \\
Dipartimento di Fisica dell'Universit\`a di Milano \\
Via Celoria 16, I-20133 Milano and \\
INFN, Sezione di Milano, Via Celoria 16, I-20133 Milano. \\
E-mail: \email{marco.caldarelli@mi.infn.it,
               dietmar.klemm@mi.infn.it,
               emanuele.zorzan@gmail.com}}
\preprint{IFUM-876-FT\\
hep-th/0610126
}

\abstract{We obtain the gravitational and electromagnetic field of
a spinning radiation beam-pulse (a gyraton) in minimal five-dimensional
gauged supergravity and show under which conditions the solution preserves
part of the supersymmetry. The configurations represent generalizations
of Lobatchevski waves on AdS with nonzero angular momentum, and possess
a Siklos-Virasoro reparametrization invariance. We compute the
holographic stress-energy tensor of the solutions and show that it transforms
without anomaly under these reparametrizations. Furthermore, we present
supersymmetric gyratons both in gauged and ungauged five-dimensional
supergravity coupled to an arbitrary number of vector supermultiplets,
which include gyratons on domain walls.
}


\begin{document}

\section{Introduction}
\label{intro}
Supersymmetry plays an important role in theoretical physics, because
solutions preserving part of the supersymmetry of the theory under
consideration benefit from good perturbative properties which permit the probing
of non-perturbative effects thanks to non-renormalization theorems. These
remarkable properties allowed for example a microscopic explanation of black hole
entropy in string theory \cite{Strominger:1996sh}.
For this reason, much effort has been dedicated to finding and classifying all
supersymmetric solutions to various supergravity theories, first using the
mathematical concept of G-structures \cite{Gauntlett:2002sc} and more recently
applying the more efficient spinorial geometry techniques \cite{Gillard:2004xq}.
This task has been successfully carried out for several supergravity theories in
diverse dimensions, cf.~e.~g.~\cite{Gauntlett:2002nw,Gauntlett:2002fz,
Gauntlett:2003fk,Caldarelli:2003pb}. However, the results
are rather implicit, since the construction of the full geometry requires the
solution of systems of differential equations that can be fairly complicated.
This makes it difficult to explore the physics behind these large classes of
supersymmetric solutions, but trying to investigate particular solutions can be
rewarding, as shown by the discovery of five dimensional supersymmetric
anti-de~Sitter black holes \cite{Gutowski:2004ez,Gutowski:2004yv},
supersymmetric black rings \cite{Elvang:2004rt}, supersymmetric G\"odel
universes \cite{Harmark:2003ud,Caldarelli:2003wh}, the LLM
solution \cite{Lin:2004nb} and many others, all obtained by detailed analysis
of the general BPS solutions to the corresponding supergravity theory.

In this paper we shall focus on both ungauged and gauged matter-coupled
supergravity in five dimensions. The latter theory plays a central role in
the AdS/CFT correspondence \cite{Aharony:1999ti}, since its asymptotically
AdS solutions are dual to four dimensional field theories. The general
supersymmetric solutions to these theories are
known \cite{Gutowski:2004yv,Gutowski:2005id}, and we shall extract a particular
subclass of geometries having the interpretation of the gravitational field
generated by gyratons in flat space or in AdS. 

Gyratons are ultrarelativistic pulsed beams of finite duration and finite
cross-section, carrying a finite amount of energy and angular momentum \cite{Frolov:2005in,Frolov:2005zq}. The
gravitational field generated by such sources is important in the study of mini
black hole production in colliders or cosmic ray experiments. Usually, the
corresponding amplitudes are computed using the Aichelburg-Sexl metric, which is
the gravitational field generated by spinless ultrarelativistic particles (see for example \cite{Kanti:2004nr} for a general review on the subject). However,
spin-spin and spin-orbit interactions can be important in certain regimes, and the
full gravitational field generated by the gyratons is important to study the
effect of the spin on the interaction of ultrarelativistic particles with spin \cite{Frolov:2005in}.

An important property of these gyraton solutions in the gauged theories is that all
curvature invariants that can be built from the metric are independent of the
functions $\Phi$ and $\Psi$ that characterize the geometry, and assume the same
value as in pure AdS space. Therefore, the spacetimes (\ref{gyratonmetric}) are
free of curvature singularities, and regular everywhere. This implies that these
solutions do not get any $\alpha'$ correction \cite{Frolov:2005ww} and are
perturbatively exact in string theory.

In this paper, we shall first find the general form of the gyraton solutions in
minimal supergravities in five dimensions, and the conditions they must satisfy
in order to preserve some supersymmetry. We then generalize these solutions
by including general vector multiplets. We obtain the supersymmetric gyratonic
solutions of the ungauged supergravities coupled to vector multiplets, and work
out the case of the $STU$ model. Then, by taking the $U(1)$ truncation of the latter
theory we recover the result of the minimal case. We study then supersymmetric
gyratons in the context of gauged supergravities coupled to an arbitrary number of
vector multiplets. In this case the equations are more complicated, but in the
special case of the gauged $STU$ model explicit supersymmetric gyraton solutions
with nonconstant scalar fields are obtained. Finally, we
compute the holographic stress tensor associated to the AdS gyratons and show
that it transforms without anomaly under the Siklos-Virasoro transformations that
leave the form of the metric invariant.

\section{Supersymmetric gyratons in minimal gauged supergravity}

Following \cite{Frolov:2005ww}, the geometry of five-dimensional AdS gyratons
can be put in the form\footnote{Our notation and conventions are as follows:
The signature is mostly minus.
The five-dimensional geometries are described by the coordinates
$(u,v,x^1,x^2,x^3)$, where $x^1 = z$, $x^2 = x$, $x^3 = y$. Late greek letters
$\mu,\nu,\ldots$ denote curved indices in five dimensions.
We will often consider lower-dimensional sections of this geometry; latin letters
$i,j,\ldots$ are indices on the three-dimensional flat space
parametrized by $(x^1,x^2,x^3)$. $\Delta^{(3)} = \partial_i\partial_i$ is the
flat Laplacian, $\hat d$ the external derivative and $\star_3$ the corresponding
Hodge operator.
Early greek letters $\al,\be,\ldots$ are indices of the two-dimensional
space $(x^2,x^3)$, again with flat metric. The antisymmetric tensor
$\varepsilon_{\al\be}$ on this space is defined such that $\varepsilon_{23}=1$, and
$\Delta^{(2)}$ is the flat Laplacian in two dimensions.}
\begin{equation}
ds^2=-\frac{l^2}{z^2}\left(
-2\,du\,dv+\Phi\,du^2
+2\left(\Psi_x dx+\Psi_y dy+\Psi_z dz\right)du
+dx^2+dy^2+dz^2
\right)\,,
\label{gyratonmetric}\end{equation}
where the functions $\Phi$ and $\Psi_i$ depend on $(u,x,y,z)$, but not on $v$.
We use this ansatz for the metric in minimal gauged supergravity, whose action
reads
\eq
S=-\frac1{16\pi G_5}\int\left(R-2\Lambda+F^2+\frac2{3\sqrt3}
\epsilon^{\mu\nu\rho\si\tau}F_{\mu\nu}F_{\rho\si}A_\tau\right)\sqrt g\,d^5x,
\label{minimalaction}\eeq
where the cosmological constant $\Lambda$ is linked to the minimal gauge
coupling $\chi$ and the curvature radius $l$ of AdS by the relations
\eq
\Lambda=\frac{\chi^2}2=\frac6{l^2}\,.
\eeq
Then, with the additional ansatz $A=A_u(u,x,y,z)\dd u$ for the gauge field, the
Maxwell equations become
\begin{equation}
\Delta^{(3)}A_u-\frac1zA_{u,z}=0\,.
\label{maxwell}\end{equation}
Note that this is just the Laplace equation on the hyperbolic space
H$^3$ with metric
\begin{displaymath}
ds_3^2 = \frac{dx^2 + dy^2 + dz^2}{z^2}\,.
\end{displaymath}
If we define $\Om$ as the curvature tensor associated to the
Sagnac connection $\Psi$,
\eq
\Omega_{ij}=\Psi_{j,i}-\Psi_{i,j}\,,
\label{defomega}\eeq
the $(ui)$ components of Einstein's equations read
\eq
\Omega_{ix,x}+\Omega_{iy,y}+\Omega_{iz,z}=\frac3z\Omega_{iz}\,,
\label{einsteinui}\end{equation}
while their $(uu)$ component, since the $R_{uu}$ component of the Ricci tensor is given by
\eq
R_{uu}=-\frac12\Delta^{(3)}\Phi+\frac3{2z}\Phi_{,z}-\frac4{z^2}\Phi+\p_u\left(\Psi_{i,i}-\frac3z\Psi_z\right)+\frac14\Omega_{ij}\Omega_{ij}\,,
\label{Ruu}\eeq
imposes an equation for $\Phi$,
\begin{equation}
\Delta^{(3)}\Phi-\frac12\Omega_{ij}\Omega_{ij}-2\partial_u\Psi_{i,i}
-\frac3{z}\left(\Phi_{,z}-2\Psi_{z,u}\right)=-\frac{4z^2}{l^2}A_{u,i}A_{u,i}
\label{siklos}\end{equation}
If equations (\ref{maxwell}), (\ref{einsteinui}) and (\ref{siklos}) are satisfied,
then the fields represent a solution of minimal gauged supergravity in five
dimensions describing a gyraton propagating in anti-de~Sitter space.

Defining ${\hat \Om} = z^{-3}\Om$, (\ref{einsteinui}) can be rewritten as
\eq
\partial_j {\hat \Om}_{ij} = 0\,,
\eeq
or equivalently $\hat d\star_3 {\hat \Om} = 0$, which implies that locally
$\star_3 {\hat \Om} = \hat d \psi$, where $\psi$ is some function of
the coordinates $x^i$ and $u$. In components we have 
\eq
\Omega_{z\al}=-2\sqrt{3}\,\frac zl\varepsilon_{\al\be}{\tilde\psi}_{,\be}\,,\qquad
\Omega_{23}=-2\sqrt{3}\,\frac zl\left({\tilde\psi}_{,z}-\frac2z
{\tilde\psi}\right)\,,
\label{omegapot}
\eeq
where we defined $\tilde\psi = -z^2 l\psi/(2\sqrt 3)$ for later convenience.
Note that (\ref{defomega}) implies that the potential $\tilde\psi$ must satisfy
equ.~(\ref{maxwell}), and thus is not completely arbitrary.

Actually, the ansatz on the gauge field is too restrictive. Starting with an
arbitrary $A_\mu$ with no dependence on $v$ to respect the symmetry generated by $\p_v$ and choosing the gauge in which $A_v=0$, it follows immediately
from Einstein's equations that $F_{zx}=F_{zy}=0$. With some additional work
one finds then that the gauge field must be of the form
\eq
A=A_u(u,x,y,z)\,\dd u+\epsilon_{\al\be}\p_\be p(u,x,y)\,\dd x^\al+A_z(u,z)\,\dd z,
\label{gengauge}\eeq
where $p(u,x,y)$ is a harmonic function on the two-space $(x,y)$,
\eq
\Delta^{(2)}p=0.
\eeq
Then the general solution is given by solving the Maxwell equation
\eq
\Delta^{(3)}A_u+\frac1z\left(\p_uA_z-\p_zA_u\right)-\p_u\p_zA_z=0,
\label{mgu}\eeq
and the Einstein equations, whose $(ui)$ components, given in
equ.~(\ref{einsteinui}) remain unchanged, while its $(uu)$ component gets some
new source from the additional components in the gauge field strength,
\begin{eqnarray}
&&\displaystyle\qquad\Delta^{(3)}\Phi-\frac12\Omega_{ij}\Omega_{ij}-2\partial_u\Psi_{i,i}
-\frac3{z}\left(\Phi_{,z}-2\Psi_{z,u}\right)\\
&=&\displaystyle
-\frac{4z^2}{l^2}\left[
\left(\p_u\p_yp-\p_xA_u\right)^2
+\left(\p_u\p_xp+\p_yA_u\right)^2
+\left(\p_uA_z-\p_zA_u\right)^2
\right].
\end{eqnarray}

According to the analysis carried out in \cite{Frolov:2005ww}, such a solution
describes a gyraton propagating in an asymptotically AdS$_5$ space-time.

Note that, if all functions $\Psi_i$ vanish, then the metric reduces to the
metric of travelling waves on AdS$_5$ (Lobatchevski waves), and equation
(\ref{siklos}) reduces to Siklos' equation \cite{Siklos:1985}. It is interesting
to note that all solutions of the supergravity equations of the travelling waves
form are supersymmetric \cite{Brecher:2000pa}. In the following, we shall
determine which of the gyraton solutions preserve some supersymmetry.

We recall that supersymmetric solutions to supergravity theories are divided
into timelike and null solutions, according to the nature of the Killing
vector constructed as a bilinear from the Killing spinor. Solutions in the null
class are all plane-fronted waves, and the gyratons are included among them.
The general null solution has been obtained in \cite{Gauntlett:2003fk} and
reads\footnote{The function $S$ appearing in \cite{Gauntlett:2003fk} is linked
to our function $\phi$ by the relation $S=e^{3\phi/2}$.}
\begin{equation}
ds^2=H^{-1}\left(\FF du^2+2\,du\,dv\right)
-H^2\left[\left(dx^1+a_1 du\right)^2 + e^{3\phi}
\left( dx^\al+e^{-3\phi} a_\al du \right)^2
\right]\,,
\label{nullmetric}\end{equation}
\eq
A=A_u\dd u+\frac{l\sqrt{3}}{4}\varepsilon_{\al\be}
\partial_{\al}\phi\,\dd x^\be\,.
\label{null sol}\end{equation}
The function $\phi(u,x^i)$ is determined by the equation
\begin{equation}
e^{2\phi} \partial^2_1 e^\phi +\Delta^{(2)}\phi=0\,.
\label{eqphi}\end{equation}
Given a solution of (\ref{eqphi}), $H(u,x^i)$ is obtained from
\begin{equation} \label{eq H}
H=-\frac{l}{2} \partial_1 \phi\,,
\end{equation}
and $A_u(u,x^i)$ is found by solving the Maxwell equation
\begin{equation}
\partial_1\left[H^2e^{2\phi}\p_1\left(e^\phi A_u\right)
\right]+\partial_\al \left(H^2\partial_\al
A_u\right)=\frac{l\sqrt{3}}{2}H\varepsilon_{\al\be}\,\partial_\al
\partial_u\phi\,\partial_\be H\,.
\label{maxwellnull}\end{equation}
Then the functions $a_i(u,x^i)$ are determined by the system
\begin{eqnarray}
\frac{1}{2 \sqrt{3}} \varepsilon_{\al\be}\p_\al\left(
H^3a_\be\right)&=&-H^2 e^{2\phi}\p_1\left(e^\phi A_u\right)\,,\nonumber\\
\frac{1}{2\sqrt3} \left[\p_\al\left(H^3 a_1\right) -
\partial_1 \left(H^3 a_\al\right)\right] &=&H^2 \varepsilon_{\al\be}
\p_\be A_u - \frac{l\sqrt3}{4}H^2\p_\al\p_u\phi\,,
\label{eqa}\end{eqnarray}
whose integrability condition is (\ref{maxwellnull}).
Finally, the function $\mathcal{F}(u,x^i)$ follows from the
$uu$-component of the Einstein equations,
\begin{equation}
R_{uu}=-2F_{u\sigma}F_u^{\
\sigma}+\frac{1}{3}g_{uu}\left(F^2+2\Lambda\right)\,.
\label{ruu}\end{equation}

To find the subset of supersymmetric null solutions of the gyraton form, we introduce the new coordinate $z$ defined by
\eq
x^1=l^3/(2z^2)
\label{defz}\eeq
and choose
\begin{equation}
\phi(z)=\ln\left(\frac{l^2}{z^2}\right)\,,
\end{equation}
as solution of (\ref{eqphi}).
Then, the general metric (\ref{nullmetric}) takes precisely the gyraton form\footnote{We also have to take $v\mapsto-v$.} (\ref{gyratonmetric}), where the functions $\Phi(u,x^i)$ and $\Psi_i(u, x^j)$ are related to the functions $\FF(u,x^i)$ and $a^i(u,x^j)$ through
\eq
\Phi=\FF+\Psi_i \Psi_i\,,\qquad
\Psi_1=-\frac{z^3}{l^3} a_1\,,\qquad
\Psi_2=\frac{z^6}{l^6}a_2\,,\qquad
\Psi_3=\frac{z^6}{l^6}a_3\,.
\eeq
With this ansatz, the Maxwell equation (\ref{maxwellnull}) for $A_u$ reduces to
(\ref{maxwell}), the equation (\ref{ruu}) for $\Phi$ simplifies to (\ref{siklos})
while the system (\ref{eqa}) for $\Psi_i$ yields the relations
\eq
\Omega_{z\al}=-2\sqrt{3}\,\frac zl\varepsilon_{\al\be}A_{u,\be}\,,\qquad
\Omega_{23}=-2\sqrt{3}\,\frac zl\left(A_{u,z}-\frac2zA_u\right)\,.
\label{omegasusy}\eeq
Observe that the equations (\ref{omegasusy}) imply (\ref{einsteinui}), and the
supersymmetric configuration we constructed represents therefore a supersymmetric gyraton propagating in anti-de~Sitter space. The converse is however not true, and not every gyratonic solution is supersymmetric. Comparing (\ref{omegasusy}) with (\ref{omegapot}) we see that the gyraton is supersymmetric if the potential $\tilde\psi$ coincides with the gauge field $A_u$. This is to be opposed to the case of the Siklos solutions ($\Psi_i=0$) which are all automatically supersymmetric \cite{Brecher:2000pa}.

An important property of the full family of gyraton solutions is that it enjoys
a large reparametrization invariance, called Siklos-Virasoro invariance.
Indeed, if we perform the diffeomorphism
\eq
\bar{u}=\chi(u)\,,\qquad
\bar v=v+\frac{\chi''}{4\chi'}x_ix_i+\lambda(u,x^j)\,,\qquad
\bar{x}^i=\sqrt{\chi'}x^i\,, \label{siklosvira}
\eeq
defined by two arbitrary functions $\chi(u)$ and $\lambda(u,x^i)$, the metric and the field equations remain invariant in form if the functions $\Phi$, $\Psi_i$ and $A_u$ transform according to
\eq
\bar{\Phi}=\frac1{\chi'}\left[\Phi+\frac12\{\chi;u\}x_i x_i-
\frac{\chi''}{\chi'}\left(\Psi_i+\lambda_{,i}\right)x_i+2\lambda_{,u}\right]\,,
\label{phitrans}
\eeq
\eq
\bar{\Psi}_i=\frac{1}{\sqrt{\chi'}}\left(\Psi_i+\la_{,i}\right)\,,
\label{psitrans}
\eeq
\eq
\bar{A}_{\overline{u}}=\frac{1}{\chi'}A_u\,,
\eeq
where
\eq
\{\chi(u);u\}=\frac{\chi'''(u)}{\chi'(u)}
-\frac32\left(\frac{\chi''(u)}{\chi'(u)}\right)^2
\eeq
denotes the Schwarzian derivative.
For Lobatchevski waves, when $\Psi_i=0$, this invariance was first obtained by
Siklos\footnote{In the original paper by Siklos there is a typeset error, corrected in \cite{Cacciatori:2004rt}.}\cite{Siklos:1985}.
Notice from (\ref{psitrans}) that the $\lambda$-transformation (with $\chi(u)=u$) acts as a gauge transformation on the connection $\Psi_i$, leaving its field strength $\Om_{ij}$ unchanged. Therefore, if $\Om_{ij}=0$ we can always perform (at least locally) a diffeomorphism to eliminate the connection $\Psi_i$ from the metric, and the solution is isometric to a Lobatchevski wave. The solution represents a genuine gyraton -- and carries angular momentum -- if and only if $\Om\neq0$.

A particular solution of (\ref{maxwell}) is given by $A_u = b(u,x^{\alpha})z^2$,
with $b(u,x^{\alpha})$ harmonic in $x^{\alpha}$, $\Delta^{(2)}b(u,x^{\alpha}) = 0$.
In the gauge $\Psi_z = 0$, which can always be achieved by a suitable
$\lambda$-transformation (\ref{psitrans}), the equations (\ref{omegasusy}) yield
for the Sagnac connection
\eq
\Psi_{\alpha} = -\frac{\sqrt 3 z^4}{2l}\epsilon_{\alpha\beta}\p_{\beta}b\,.
\eeq
Finally, the Siklos equation (\ref{siklos}) is solved by
\eq
\Phi = \alpha(u,x,y)z^4 + \beta(u,x,y)z^6\,,
\eeq
where the functions $\alpha$ and $\beta$ obey
\begin{eqnarray}
&& \Delta^{(2)}\beta - \frac 8{l^2}\p_{\alpha} b\p_{\alpha} b = 0\,, \nonumber \\
&& \Delta^{(2)}\alpha + 12\beta + \frac{16}{l^2}b^2 = 0\,.
\end{eqnarray}
In the special case $b=\beta=0$, $\alpha =$ const., this solution reduces to
the five-dimensional generalization of the Kaigorodov spacetime \cite{kaigorodov},
constructed in \cite{Cvetic:1998jf}.

Finally, let us consider the ungauged limit. To obtain a finite result we first have to introduce the new coordinate $\eta$ defined by
\eq
\frac zl=\exp\left(\frac\eta l\right)\,,
\eeq
then the $l\rightarrow\infty$ limit can be performed safely. We obtain the geometry of plane-fronted waves with parallel rays (pp-waves), whose metric is
\eq
ds^2=\FF\,du^2+2du\,dv-\left(dx^i+\Psi_i\,du\right)^2\,,
\label{umetric}\eeq
and the gauge potential reads
\eq
A=-\frac1{2\sqrt3}\varphi(u,x^i)\dd u\,,
\label{defphi}\eeq
where we have defined $\FF=\Psi_i\Psi_i-\Phi$ and $\varphi=-2\sqrt3A_u$.
Then, Maxwell's equations (\ref{mgu}) imply that $\varphi$ has to be a harmonic
function on the flat three-dimensional base space, $\Delta^{(3)}\varphi=0$,
while the $(ui)$ components of the Einstein equations are solved iff the
curl of $\vec\Psi$ is the gradient of some function $\psi(u,x^i)$ on $\R^3$,
\eq
\vec\nabla\wedge\vec\Psi=\vec\nabla\psi.
\label{ua}\eeq
Finally, the $(uu)$ component of Einstein's equations reads
\eq
\Delta^{(3)}\FF+2\D\vec\nabla\cdot\vec\Psi-2\p_{(i}\Psi_{j)}\p_{(i}\Psi_{j)}=\frac13\left(\nabla\varphi\right)^2\,,
\label{uuu}\eeq
where we have defined the operator $\D$ acting on scalar functions $f$ as
$\D f=\p_uf-\Psi^i\p_if$.
Therefore, one obtains a solution of minimal ungauged supergravity with a metric
of the form (\ref{umetric}) by choosing a function $\varphi(u,x^i)$ harmonic on
$\R^3$ and an arbitrary function $\psi(u,x^i)$, and then solving (\ref{uuu})
for $\FF(u,x^i)$. It is easy to show that this is the most general solution
of minimal supergravity with metric of the form (\ref{umetric}) up to gauge
transformations\footnote{The functions $p(u,x,y)$ and $A_z(u,z)$ appearing in
(\ref{gengauge}) can indeed be gauged away in the $l\to\infty$ limit.}.
The supersymmetry condition then simply requires the equality of the functions
$\varphi$ and $\psi$, $\varphi=\psi$.
This condition can be equivalently obtained by taking the ungauged limit of
equations (\ref{omegasusy}).

Observe that we are free to impose the gauge condition
$\vec\nabla\cdot\vec\Psi=0$ by performing a diffeomorphism; then the equation
for $\FF$ simplifies and reads
\eq
\Delta^{(3)}\FF=2\p_{(i}\Psi_{j)}\p_{(i}\Psi_{j)}+\frac13\left(\nabla\varphi\right)^2\,.
\label{uf}\eeq
This family of supersymmetric solutions to minimal supergravity in five
dimensions was already obtained in \cite{Gauntlett:2002nw}, as part of the
general supersymmetric configuration in the null class, and generalized in the
non-supersymmetric case to arbitrary dimension in \cite{Frolov:2005ja}.


\section{Inclusion of vector multiplets}
The general supersymmetric solution for $\NN=2$, $D=5$ supergravity, coupled to
$n-1$ abelian vector multiplets was obtained in \cite{Gutowski:2005id}, both for
the ungauged and the gauged theories. The action for these supergravity theories
is given by \cite{Gunaydin:1983bi,Gunaydin:1984ak}
\begin{eqnarray}
S=-\frac1{16\pi G}\int\left[\left(R-2\chi^2 V\right)\star{\mathbf 1}
+Q_{IJ}F^I\w\star F^J\phantom{\frac12}\right.\qquad\qquad\quad\nonumber\\
\left.-Q_{IJ}\,\dd X^I\w\star\dd X^J+\frac16C_{IJK}F^I\w F^J\w A^K\right]\,.
\end{eqnarray}
Here the indices $I, J, K,\ldots=1,\ldots,n$ label the vector multiplets,
$A^I$ are the $n$ $U(1)$ gauge fields and $F^I$ denote the corresponding field
strengths. The graviphoton of the theory is given by the linear combination of the
gauge fields
${\mathcal A}=V_IA^I$, determined by the constant coefficients $V_I$.
The scalars $X^I$ are constrained by the relation
\eq
\V = \frac 16 C_{IJK}X^IX^JX^K = 1\,,
\label{scalarconstraint}\eeq
so that one has $n-1$ independent scalars. The homogeneous cubic polynomial
$\V$ defines a "very special geometry" \cite{deWit:1992cr}. The
coefficients $C_{IJK}$ are constants that are symmetric in $I,J,K$,
and can be used to define the fields $X_I$ with lower indices,
\eq
X_I=\frac16C_{IJK}X^JX^K\,,
\eeq
in terms of which the scalar constraint becomes $X_I X^I=1$.
The metric $Q_{IJ}$ depends on the $X^I$ via
\eq
Q_{IJ}=\frac92X_IX_J-\frac12C_{IJK}X^K\,.
\label{STUQ}\eeq
We will assume it is invertible, and denote its inverse by $Q^{IJ}$. The
following relations are then shown to hold,
\eq
Q_{IJ}X^J = \frac32 X_I\,,\qquad
Q_{IJ}\p_{\mu} X^J = -\frac32 \p_{\mu} X_I\,.
\eeq
Finally, the potential for the scalar fields is given by
\eq
V = 9V_IV_J\left(X^IX^J-\frac12Q^{IJ}\right)\,.
\eeq
For Calabi-Yau compactifications of M-theory, $\V$ denotes the intersection form,
$X^I$ and $X_I$ correspond to the size of the two- and four-cycles 
of the Calabi-Yau threefold respectively, and $C_{IJK}$ are the intersection
numbers of the threefold. In this case, $n-1$ is given by the Hodge number
$h_{(1,1)}$.

The solutions of these models in which only the photon and graviton are excited
solve as well the minimal model (\ref{minimalaction}). These are the subclass of
solutions with constant scalar fields and all vector fields proportional to the
graviphoton. Indeed, setting
\eq
X_I=\Xi V_I\,,\qquad A^I=\frac2{\sqrt3}X^IA\,,
\label{gaugeminimal}\eeq
we recover the minimal gauged supergravity action (\ref{minimalaction}), with the
normalization $A=(\sqrt3\Xi/2){\mathcal A}$ for the graviphoton and an effective
gauge coupling constant $\chi_{\mathrm{min}}$ given by
\eq
\chi^2_{\mathrm{min}}=\frac{12\chi^2}{\Xi^2}\,.
\eeq
On the other hand, ungauged $N=2$ supergravities coupled to vector multiplets
are obtained by taking the limit $\chi \to 0$.

Of particular interest are the $STU$ models, which represent truncations
of the dimensional reduction of higher-dimensional supergravities coming from
string theory. These models, in both the ungauged and gauged case, are obtained
for $n=3$ by taking $C_{123}$ and its permutations equal to one and
otherwise zero. Furthermore one has $V_I = 1/3$ in these models. Then
\eq
Q_{IJ}=\frac1{2(X^I)^2}\delta_{IJ}\,,
\eeq
$X_I=1/(3X^I)$, and the scalar potential reads
\eq
V = 2\left(\frac 1{X^1} + \frac 1{X^2} + \frac 1{X^3}\right)\,.
\eeq
The minimal supergravities of the previous section can then be obtained as
truncations of the $STU$ models using the prescription (\ref{gaugeminimal}),
for instance by taking $\Xi=1$, all gauge fields equal and normalized as in
(\ref{gaugeminimal}), and constant scalar fields $X^I=1$.
Moreover, one also needs to rescale the gauge coupling,
$\chi^2_{\mathrm{min}}=12\chi^2$.

In the following, we shall find supersymmetric gyraton solutions to these
supergravity theories.

\subsection{Ungauged supergravity coupled to vector multiplets}

We first consider the ungauged model, and subsequently generalize the results to
the gauged case. The general null solution in the ungauged supergravity has a
geometry described by the metric \cite{Gutowski:2005id}
\eq
ds^2=H^{-1}\left(\FF\,du^2+2du\,dv\right)-H^2\left(dx^i+a_idu\right)^2,
\eeq
and the gauge fields are of the form
\eq
F^I=\dd u\w Y^I+\star_3\hat\dd(HX^I)\,,
\label{F}\eeq
where $Y^I(u,x^i)$ are $n$ one-forms on the base space $\R^3$. The scalars $X^I$
are given in terms of $n$ arbitrary $u$-dependent harmonic functions $K^I(u,x^i)$
on $\R^3$,
\eq
X^I=H^{-1}K^I\,.
\eeq
The scalar constraint (\ref{scalarconstraint}) translates therefore into the relation
\eq
H^3=\frac16C_{IMN}K^IK^MK^N\,,
\eeq
which determines the function $H(u,x^i)$.
The one-forms $Y^I$ on $\R^3$ have to satisfy the constraint
\eq
X_IY^I=\frac13H^{-2}\star_3\hat\dd\left(H^3a\right),
\label{uvaa}\eeq
the Bianchi identity
\eq
\hat\dd Y^I=\star_3\hat\dd\left(\p_uK^I\right)\,,
\label{uvbianchi}
\eeq
and the Maxwell equations
\eq
C_{IJK}\left(Y^J{}_i\nabla^iK^K+\frac12\nabla^iY^J{}_iK^K\right)=0.
\eeq
Finally, the function $\FF(u,x^i)$ is determined by the $(uu)$ component of
Einstein's equations,
\eq
\Delta^{(3)}\FF=-2H\left(\D\hat W-W_{ij}W^{ij}\right)
+2HQ_{IJ}\left(U^I{}_iU^{Ji}+H^2\D X^I\D X^J\right)\,,
\eeq
where we have defined
\eq
W_{ij}=H\p_{(i}a_{j)}-\D H\delta_{ij}\,,\qquad \hat W=W^i{}_i\,,
\eeq
\eq
U^I=Y^I+\star_3\left[a\w\hat\dd\left(HX^I\right)\right]\,,
\eeq
and the operator $\D$ acts on scalar functions $f$ as $\D f=\p_uf-a^i\p_if$.

A plane-fronted wave is said to be a plane-fronted wave with parallel rays
(pp-wave) if $\dd u$ is covariantly constant. This condition imposes without
loss of generality $H=1$. We also make the ansatz $A^I=A^I{}_u\dd u$ for the
gauge fields, by analogy with the minimal case, where the supersymmetric gyratons
have only this component of the gauge field\footnote{In presence of vector
multiplets, this is not the most general supersymmetric solution of the gyratonic
form.}. Then, comparing with (\ref{F}), we deduce that $Y^I=-\p_iA^I{}_u\,\dd x^i$
and that $X^I$, and therefore $K^I$, are functions of $u$ only. The Bianchi
identity (\ref{uvbianchi}) is then automatically satisfied, the Maxwell equations
reduce to
\eq
C_{IMN}K^M\Delta^{(3)}A^N{}_u=0\,,
\eeq
and the connection $a_i$ is determined by equation (\ref{uvaa}).

To summarize, the generic supersymmetric gyraton is obtained by choosing $n$
arbitrary functions $X^I=X^I(u)$ subject to the constraint $C_{IJK}X^IX^JX^K=6$.
Then the gauge potentials are determined by
\eq
C_{IJK}X^J\Delta^{(3)}A^K{}_u=0\,,
\label{uvmax}\eeq
and the connection $a_i$ by
\eq
\ep_{ijk}\p_ja_k=-\frac12C_{IJK}X^JX^K\p_iA^I{}_u\,.
\label{uva}\eeq
The equation for $\FF$ follows then from the $(uu)$ component of Einstein's
equations,
\eq
\Delta^{(3)}\FF=-2\left(\D\p_ia_i-\p_{(i}a_{j)}\p_{(i}a_{j)}\right)+
2Q_{IJ}\left(\p_iA^I{}_u\p_iA^J{}_u+\p_uX^I\p_uX^J\right).
\label{uvF}
\eeq
Note that, by an appropriate diffeomorphism, we can always choose coordinates
such that $\vec\nabla\cdot\vec a=0$, so that (\ref{uvF}) simplifies to
\eq
\Delta^{(3)}\FF=2\p_{(i}a_{j)}\p_{(i}a_{j)}+
2Q_{IJ}\left(\p_iA^I{}_u\p_iA^J{}_u+\p_uX^I\p_uX^J\right).
\eeq
More progress can be done in the $STU$ model. Recalling $n=3$ and $C_{123}=1$
is the only non-vanishing component of $C_{IJK}$, up to permutations, the solution
is given by three arbitrary functions $X^I(u)$ subject to the constraint
$X^1X^2X^3=1$ and three $u$-dependent harmonic functions $A^I{}_u(u,x^i)$ on the
flat base space $\R^3$, $\Delta^{(3)}A^I{}_u=0$. Finally, the connection $a_i$ is
determined by the equation
\eq
\nabla\w a=-\nabla\sum_{I=1}^3 (X^I)^{-1}A^I{}_u\,.
\label{STUa}\eeq
Using (\ref{STUQ}) the equation for $\FF$ becomes (in the $\vec\nabla\cdot a=0$
gauge)
\eq
\Delta^{(3)}\FF=2\p_{(i}a_{j)}\p_{(i}a_{j)}+
\sum_{I=1}^3\frac1{(X^I)^2}\left[\left(\nabla A^I{}_u\right)^2+
\left(\p_uX^I\right)^2\right].
\label{STUf}\eeq
The minimal supergravity gyraton of the previous section is then easily recovered
taking $X^I=1$ and all gauge fields $A^I$ equal and proportional to the graviphoton
field $A$, as in equation (\ref{gaugeminimal}).
Then, the potential $\varphi$ defined in (\ref{defphi}) is $\varphi=-3A^I{}_u$ for any
$I$. With these definitions, equations (\ref{STUa}) and (\ref{STUf}) reduce
respectively to (\ref{ua}) and (\ref{uf}), and the complete solution corresponds
to a gyraton of minimal ungauged supergravity.

\subsection{Gauged supergravity coupled to vector multiplets}

Let us turn now to the gauged case. The general null solution is given by the
metric \cite{Gutowski:2005id}
\begin{eqnarray}
ds^2=H^{-1}\left(\FF\,du^2+2du\,dv\right)-H^2\left[\left(dx^1+a_1du\right)^2
\right.\qquad\qquad\nonumber\\
\left.+\left(S\,dx^2+S^{-1}a_2du\right)^2+\left(S\,dx^3+S^{-1}a_3du\right)^2\right],
\end{eqnarray}
and the gauge fields are of the form
\eq
F^I=\dd u\w Y^I+\star_3\left[\hat\dd(HX^I)-3\chi H^2Q^{IJ}V_J\dd x^1\right]\,,
\label{gF}\eeq
where $Y^I(u,x^i)$ are $n$ one-forms on the three-dimensional base space
\eq
ds_3^2=\left(dx^1\right)^2+S^2\left[\left(dx^2\right)^2+\left(dx^3\right)^2\right]\,,
\eeq
whose metric tensor will be denoted by $h_{ij}$. Covariant derivatives with respect
to this metric are indicated by $\hat\nabla_i$, and external derivatives by
$\hat\dd$ and from now on $\star_3$ denotes the Hodge dual on the curved base space.
The corresponding gauge potentials $A^I$ have to satisfy the constraint
\eq
V_IA^I=V_IA^I{}_u\,\dd u+\frac1{3\chi}S^{-1}\left(\p_2S\,\dd x^3-\p_3 S\,
\dd x^2\right).
\label{gVA}\eeq
The functions $S$ and $H$ are linked by
\begin{equation}
\p_1S=-3\chi V_IX^ISH\,,
\label{eqH}\end{equation}
\begin{equation}
\square\ln S=-9\chi^2H^2\left(Q^{IJ}-2X^IX^J\right)V_IV_J=2\chi^2H^2 V\,.
\label{eqS}\end{equation}
Here, $\square$ is the Laplace operator on the three-dimensional base space $ds^2_3$.
The one-forms $Y^I$ have to satisfy the constraint
\eq
X_IY^I=\frac13H^{-2}\star_3\hat\dd\left(H^3a\right)-2\chi HV_JA^J{}_u\dd x^1,
\label{va}\eeq
which yield an equation for $a^i$, while the Bianchi identity imposes the conditions
\eq
\hat\dd Y^I=\p_u\left[
\star_3\left(\hat\dd\left(HX^I\right)-3\chi H^2Q^{IJ}V_J\,\dd x^1\right)\right]\,,
\label{vbianchi}\eeq
\eq
\hat\dd\star_3\hat\dd\left(HX^I\right)=3\chi S^{-2}\p_1\left(S^2H^2Q^{IJ}
\right)V_J\,\dd\mathrm{Vol}_3\,,
\label{eqX}\eeq
and the Maxwell equations read
\begin{eqnarray}
\hat\dd\left(HQ_{IJ}\star_3Y^J\right)=3\hat\dd\left(H^3a\right)\w
\left[\frac12\hat\dd\left(H^{-1}X_I\right)+\chi V_I\dd x^1\right]\qquad\qquad\quad
\nonumber\\
+\frac12C_{IJK}\star_3Y^J\w\left[\hat\dd\left(HX^K\right)
-3\chi H^2Q^{KP}V_P\,\dd x^1\right]\,.
\label{vmaxwell}\end{eqnarray}
Finally, the function $\FF(u,x^i)$ is determined by the $(uu)$ component of
Einstein's equations
\eq
\square\FF=-2H\left(\D\hat W-W_{ij}W^{ij}\right)
+2HQ_{IJ}\left(U^I{}_iU^{Ji}+H^2\D X^I\D X^J\right)\,,
\label{eqF}\eeq
where we have defined
\eq
W_{ij}=H\hat\nabla_{(i}a_{j)}-\left(\D H\right)h_{ij}-\frac12H\p_uh_{ij}\,,
\qquad \hat W=W^i{}_i\,,
\eeq
\eq
U^I=Y^I+\star_3\left[a\w
\left(\hat\dd\left(HX^I\right)-3\chi H^2Q^{IJ}V_J\,\dd x^1\right)
\right]\,,
\eeq
and the operator $\D$ acts on scalar functions $f$ as $\D f=\p_uf-a^i\p_if$.

To obtain a supersymmetric solution with metric of the gyratonic form in presence
of vector multiplets, we make the following ansatz for the functions $H$ and $S$,
\eq
H=-\frac l{2x^1}\,,\qquad S=\left(\frac{2x^1}l\right)^{3/2}\,,
\label{HS}\eeq
in analogy with what we found in the minimal case.
Then, equations (\ref{eqH}) and (\ref{eqS}) are satisfied if we impose the
constraints
\eq
V_IX^I=\frac1{\chi l}\,,\qquad
Q^{IJ}V_IV_J=\frac2{3\chi^2l^2} \label{constrX}
\eeq
on the scalar fields. It immediately follows that the scalar potential is
constant with value
\eq
V = \frac6{\chi^2l^2}\,,
\eeq
so it acts as an effective cosmological constant.
Moreover, the scalar equation (\ref{eqX}) can be rewritten as the vanishing of
the external derivative of a two-form, and is solved if there exist $n$ one-forms
$\psi^I(u,x^i)$ on the three-dimensional base space such that
\eq
\star_3\left[\hat\dd(HX^I)-3\chi H^2Q^{IJ}V_J\dd x^1\right]=\hat\dd\psi^I(u,x^i)\,.
\label{defpsi}\eeq
Comparing with equation (\ref{gF}), we see that $F^I=\dd u\wedge Y^I +
\hat\dd\psi^I$, while the Bianchi identity (\ref{vbianchi}) implies the
existence of functions $\lambda^I(u,x^i)$ such that
$Y^I=\p_u\psi^I-\hat\dd\lambda^I$. Plugging back in
(\ref{gF}) one discovers then that the newly introduced quantities are the
components of the gauge fields,
\eq
A^I{}_u=\lambda^I,\qquad
A^I{}_{i}\,\dd x^i=\psi^I.
\eeq
Finally, since there is no dependence on $x^\al$ in $S$, equation (\ref{gVA})
implies the vanishing of the spatial components of the graviphoton,
\eq
V_I\psi^I=0.
\eeq
The general solution can then be found by solving Maxwell's equation
(\ref{vmaxwell}) for $\lambda^I$ and $\psi^I$, equation (\ref{va}) for $a_i$
and finally (\ref{eqF}) for $\F$. Observing that in the minimal supergravities
the supersymmetric solutions have $A_i=0$, we make the simplifying ansatz
$\psi^I=0$. Then, equation (\ref{defpsi}) states the independence of the scalar
fields on the variables $x^\al$, $X^I=X^I(u,x^1)$, and its $x^1$ component imposes
\eq
\p_1X_I=-2\chi H\left(V_I-\frac{H'}{2\chi H^2}X_I\right),
\label{pX}\eeq
with $H'=\p_1H$. This can be integrated to give
\eq
X_I = \chi l V_I + \frac{\kappa_I(u)}{x^1}\,,
\eeq
where $\kappa_I(u)$ are $u$-dependent integration constants. Contracting with
$X^I$ and using the constraints (\ref{constrX}) leads to $X^I\kappa_I = 0$,
whereas contraction with $Q^{IJ}V_J$ implies $Q^{IJ}\kappa_I V_J = 0$.
Using these identities and contracting finally with $Q^{IJ}\kappa_J$ one gets
$Q^{IJ}\kappa_I\kappa_J = 0$ and thus $\kappa_I = 0$, because $Q_{IJ}$ is
positive definite. The scalars $X_I$ are thus constant. The Maxwell equations
(\ref{vmaxwell}) imply that the gauge potentials $\lambda^I = A^I{}_u$
satisfy (\ref{maxwell}), where the coordinate $z$ is defined by (\ref{defz}).
Note that this does not mean that the vector fields $A^I$ can be written as
in (\ref{gaugeminimal}), so we are not necessarily in the minimal
supergravity theory.

A more interesting solution with nonconstant scalar fields can be found
for the $STU$ model. To this end we have to relax the ansatz (\ref{HS}) and
allow for an arbitrary dependence of the functions $H$, $S$ and $X^I$ on the
coordinates $u$ and $x^1$ while keeping them independent of
$x^{\alpha}$. Then equation (\ref{eqX}) can be integrated once,
\eq
\p_1\xi^I-2\chi\left(\xi^I\right)^2=S^{-2}s^I\,,
\label{eqxi}\eeq
where we have defined for convenience $\xi^I(u,x^1)=HX^I$ and the functions $s^I(u)$ are arbitrary integration constants. It follows from (\ref{defpsi}) that
\eq
\hat\dd\psi^I=s^I(u)\,\dd x^2\w\dd x^3\,,
\eeq
and therefore $s^I(u)$ represent the magnetic fields in the transverse space. It is however difficult to solve the equations in this case, and we shall take $s^I(u)=0$ for simplicity. Then, equation (\ref{eqxi}) can be integrated,
\eq
\xi^I(u,x^1)=\frac1{c^I(u)-2\chi x^1}\,,
\eeq
where the arbitrary functions $c^I(u)$ are integration constants.
Integration of (\ref{eqH}) yields $S(u,x^1)$,
\eq
S(u,x^1)=\al(u)\prod_I\sqrt{c^I(u)-2\chi x^1}\,,
\eeq
and the condition $\xi^1\xi^2\xi^3=H^3$ gives for $H(u,x^1)$
\eq
H(u,x^1)=\prod_I\frac1{\left(c^I(u)-2\chi x^1\right)^{1/3}}\,, \label{HSTU}
\eeq
so that the scalar fields take the form
\eq
X^I(u,x^1)=\frac{\prod_J\left(c^J(u)-2\chi x^1\right)^{1/3}}{c^I(u)-2\chi x^1}\,.
\eeq
(\ref{eqS}) is then satisfied without further restrictions.
Then, Maxwell's equations (\ref{vmaxwell}) read
\eq
\p_1^2A^I{}_u-\frac{4\chi}{c^I(u)-2\chi x^1}\p_1A^I{}_u+S^{-2}\Delta^{(2)}A^I{}_u=0\,,
\label{sid}\eeq
and the functions $\Psi_i=H^3a_i$ are determined by solving the system
\begin{eqnarray}
\p_2\Psi_3-\p_3\Psi_2&=&-\al^2(u)\sum_I\left[\left(c^I-2\chi x^1\right)\p_1A^I{}_u-2\chi A^I{}_u\right]\,,\\
\p_1\Psi_\al-\p_\al\Psi_1&=&-\frac{\al^2}{S^2}\sum_I\left(c^I-2\chi x^1\right)\varepsilon_{\al\be}\p_\be A^I{}_u\,.
\end{eqnarray}
Finally, the function $\F$ is obtained by solving the Siklos equation (\ref{eqF}).

A nontrivial solution of (\ref{sid}) is given by
\eq
A^I{}_u=\frac{b^I(u,x^\al)}{c^I(u)-2\chi x^1}+a^I(u,x^\al)\,,
\eeq
where the integration constants $a^I(u,x^\al)$ and $b^I(u,\al)$ are arbitrary functions of $u$ and harmonic functions in the two-dimensional flat space $(x^2,x^3)$,
\eq
\Delta^{(2)}a^I(u,x^\al)=0\,,\qquad
\Delta^{(2)}b^I(u,x^\al)=0\,.
\eeq
Then the system of equations for $\Psi_i$ reduces to
\begin{eqnarray}
\p_2\Psi_3-\p_3\Psi_2&=&2\chi\al^2(u)\sum_Ia^I(u,x^\al)\,,\\
\p_1\Psi_\al-\p_\al\Psi_1&=&-\frac{1}{\prod_J\left(c^J-2\chi x^1\right)}\varepsilon_{\al\be}\p_\be\sum_I\left[b^I+a^I\left(c^I-2\chi x^1\right)\right]\,,
\end{eqnarray}
that can be readily integrated in the $\Psi_1=0$ gauge (which can always be
chosen by a suitable diffeomorphism \cite{Gutowski:2005id}) to give
\eq
\Psi_{\al}=-\varepsilon_{\al\be}\p_\be\sum_I\int\frac{b^I+a^I\left(c^I-2\chi x^1\right)}{\prod_J\left(c^J-2\chi x^1\right)}dx^1+f_\al(u,x^\be)\,,
\label{integral}\eeq
where the functions $f_\al(u,x^\be)$ are solutions of the equation
\eq
\p_2f_3-\p_3f_2=2\chi\al^2(u)\sum_Ia^I(u,x^\al)\,.\\
\eeq
The explicit evaluation of the integral (\ref{integral}) depends on the number
of functions $c^I(u,x^\al)$ that coincide, and we have therefore three possible
cases. If all functions $c^I$ are distinct, we have
\begin{displaymath}
\Psi_\al=\frac1{2\chi}\varepsilon_{\al\be}\p_\be\left[
\sum_{I,J}\frac{b^I\ln\left(c^J-2\chi x^1\right)}{\prod_{K\neq J}\left(c^K-c^J\right)}
-\frac12\sum_{I\neq J\neq K}\frac{a^I}{c^J-c^K}\ln\left(\frac{c^J-2\chi x^1}{c^K-2\chi x^1}\right)
\right]+f_\al\,.
\end{displaymath}
If two of them coincide, say $c^1=c^2=c$ and $c^3=\tilde c$ with $\tilde c\neq c$, then
\begin{displaymath}
\Psi_\al=-\frac1{2\chi}\varepsilon_{\al\be}\p_\be\left[
\left(\frac{\sum_I b^I}{(c-\tilde c)^2}+\frac{a^1+a^2}{c-\tilde c}
\right)
\ln\frac{c-2\chi x^1}{\tilde c-2\chi x^1}
+\left(a^3-\frac{\sum_I b^I}{(c-\tilde c)}\right)\frac1{c-2\chi x^1}
\right]+f_\al\,.
\end{displaymath}
Finally, if the three functions coincide, $c^1=c^2=c^3=c$, we obtain
\begin{displaymath}
\Psi_\al=-\frac1{4\chi\left(c-2\chi x^1\right)}\varepsilon_{\al\be}\p_\be\left[
\sum_I\left(\frac{b^I}{c-2\chi x^1}+2a^I\right)
\right]+f_\al\,.
\end{displaymath}
We finally note that in the special case $A^I{}_u = 0$, $\Psi_i = 0$,
$c^I(u) = c^I =$ const., $\alpha(u) = 1$, the metric reduces to
\begin{eqnarray}
ds^2=H^{-1}\left(\FF\,du^2+2du\,dv + \left(dx^2\right)^2 + \left(dx^3\right)^2
\right)-H^2\left(dx^1\right)^2\,, \label{domainwall}
\end{eqnarray}
where $H$ is given by (\ref{HSTU}) with constant $c^I$, and $\FF$ obeys
\eq
\p_1(H^{-3}\p_1\FF) + \Delta^{(2)}\FF = 0\,.
\eeq
The geometry (\ref{domainwall}) represents a wave propagating on a domain
wall\footnote{For $\FF = 0$, (\ref{domainwall}) reduces to a subclass of the
domain wall solutions found in \cite{Cacciatori:2003kv}.}, whereas the more general
solutions with $\Psi_i \neq 0$ presented above describe gyratons on a domain wall
background.

\section{Holographic stress tensor}

In this section we compute the holographic stress tensor \cite{Balasubramanian:1999re}
associated to the solutions (\ref{gyratonmetric}) of minimal gauged supergravity.
Let us consider the hypersurface $z=\mathrm{const}$, with unit normal vector
$u^\mu=(0,0,0,0,z/l)$, and induced metric
\begin{equation}
d\sigma^2=-\frac{l^2}{z^2}\left(
-2\,du\,dv+\left(\Phi-\Psi_z^2\right)du^2
+2\left(\Psi_x dx+\Psi_y dy\right)du
+dx^2+dy^2
\right)\,.
\end{equation}
It is convenient to introduce a new function $\zeta = \Phi-\Psi_z^2$, since $\Phi$
enters the induced metric (and the extrinsic curvature, as we will shortly see) only
through this combination.
Then, the extrinsic curvature $K_{ab}=\sigma_a{}^c\sigma_b{}^d\nabla_c u_d$
of the $z=$const hypersurface is\footnote{Early latin indices $a,b,\ldots$
are used only in this section; $x^a$ are the coordinates $u,v,x,y$.}
\begin{equation}
K_{ab}=\left(\begin{array}{cccc}
\displaystyle\frac{l}{2z^2}\left(z\zeta_{,z}-2\zeta\right)
& \displaystyle \frac l{z^2}
& \displaystyle \frac{l}{2z^2}\left(z\Psi_{x,z}-2\Psi_x\right)
& \displaystyle \frac{l}{2z^2}\left(z\Psi_{y,z}-2\Psi_y\right)\\
\\
\displaystyle \frac{l}{z^2} & 0 & 0 & 0 \\
\\
\displaystyle \frac{l}{2z^2}\left(z\Psi_{x,z}-2\Psi_x\right) & 0
& \displaystyle -\frac{l}{z^2} & 0 \\
\\
\displaystyle \frac{l}{2z^2}\left(z\Psi_{y,z}-2\Psi_y\right) & 0 & 0
& \displaystyle -\frac{l}{z^2}
\end{array}\right)\,. \label{extrcurv}
\end{equation}
Its trace is $K=K^a{}_a=4$, and the only non-vanishing components of the Einstein
tensor of the $z=$const hypersurface are
\begin{displaymath}
G_{uu}=-\frac12\Delta^{(2)}\zeta + \Psi_{x,xu}+\Psi_{y,yu}+
\frac12\left(\Omega_{xy}\right)^2\,,
\end{displaymath}
\begin{equation}
G_{ux}=\frac12\Omega_{xy,y}\,,
\qquad\qquad
G_{uy}=\frac12\Omega_{yx,x}\,. \label{Einsttens}
\end{equation}
The holographic energy-momentum tensor is given by \cite{Balasubramanian:1999re}
\begin{equation}
T_{ab} = \frac{1}{8\pi G_5}\left[
\left(K_{ab}-K\sigma_{ab}\right)+\frac 3l\sigma_{ab}+\frac l2 G_{ab}\right]\,.
\end{equation}
Using (\ref{extrcurv}) and (\ref{Einsttens}), we find that the only
non-vanishing components read
\begin{displaymath}
T_{uu} = \frac{l}{16\pi G_5}\left[\frac1{z}\zeta_{,z}-\frac12\Delta^{(2)}
         \zeta + \Psi_{x,xu}+\Psi_{y,yu}+\frac12\left(\Omega_{xy}\right)^2\right]\,,
\end{displaymath}
\begin{displaymath}
T_{ux} = \frac{l}{16\pi G_5}\left[\frac1{z}\Psi_{x,z}+\frac12\Omega_{xy,y}\right]\,,
\end{displaymath}
\begin{equation}
T_{uy} = \frac{l}{16\pi G_5}\left[\frac1{z}\Psi_{y,z}+\frac12\Omega_{yx,x}\right]\,.
\end{equation}
The stress tensor $\hat T_{ab}$ of the associated conformal field theory is
then obtained by
\begin{equation}
\hat T_{ab} = \lim_{z\rightarrow0}\frac{l^2}{z^2}T_{ab}\,,
\end{equation}
and its components are
\begin{equation}
\hat T_{uu} = \frac{N^2}{8\pi^2}\lim_{z\rightarrow0}\frac1{z^2}\left[
\frac1{z}\zeta_{,z}-\frac12\Delta^{(2)}\zeta + \Psi_{x,xu}+\Psi_{y,yu}
+\frac12\left(\Omega_{xy}\right)^2\right]\,, \label{hatTuu}
\end{equation}
\begin{equation}
\hat T_{ux} = \frac{N^2}{8\pi^2}\lim_{z\rightarrow0}\frac1{z^3}
\left(\Psi_{x,z}+\frac z2\Omega_{xy,y}\right)\,, \label{hatTux}
\end{equation}
\begin{equation}
\hat T_{uy} = \frac{N^2}{8\pi^2}\lim_{z\rightarrow0}\frac1{z^3}
\left(\Psi_{y,z}+\frac z2\Omega_{yx,x}\right)\,, \label{hatTuy}
\end{equation}
where we used the AdS/CFT dictionary $l^3/G_5 = 2N^2/\pi$. Note that the
energy-momentum tensor is traceless, $\hat T = \sigma^{ab}\hat T_{ab} = 0$,
so there is no conformal anomaly in the dual CFT.

Let us now consider the effect of a Siklos-Virasoro transformation (\ref{siklosvira})
on the components of the holographic stress tensor. To this end, we choose the
gauge $\Psi_z = 0$, which can always be achieved by a gauge transformation
(\ref{psitrans}). The transformations that preserve this gauge obey
$\partial_z \lambda = 0$. It is then straightforward to show that under
(\ref{siklosvira})
\begin{equation}
T_{uu} \to \bar T_{uu} = \frac 1{\chi'^2}T_{uu} + \frac l{32\pi G_5}
\frac{\chi''}{\chi'^3}\left(\partial_z - \frac 1z\right)x^{\alpha}
\Omega_{\alpha z}\,,
\end{equation}
\begin{displaymath}
T_{ux} \to \bar T_{ux} = \frac 1{\chi'^{3/2}}T_{ux}\,, \qquad
T_{uy} \to \bar T_{uy} = \frac 1{\chi'^{3/2}}T_{uy}\,.
\end{displaymath}
Contrary to the wave profile $\Phi$ (cf.~(\ref{phitrans})), the component
$T_{uu}$ transforms without anomaly term proportional to the Schwarzian
derivative. The reason for this is that the Schwarzian derivative coming
from the term $z^{-1}\partial_z \tilde\Phi$ cancels the one appearing in
$-\Delta^{(2)}\tilde\Phi/2$. The stress tensor proposed in \cite{Banados:1999tw}
has only the former term and transforms therefore with anomaly. As the second piece
(which stems from the Einstein tensor, i.~e.~, from the counterterms that are
necessary to cancel divergences) is absent in \cite{Banados:1999tw}, their
stress tensor is divergent in the $z\to 0$ limit.

In order to obtain a more explicit expression for the holographic energy-momentum
tensor, we use the Fefferman-Graham expansion \cite{fg},
\eq
g_{\mu\nu} = \frac{l^2}{z^2}g_{\mu\nu}^{(0)} + g_{\mu\nu}^{(2)} +
\frac{z^2}{l^2}g_{\mu\nu}^{(4)} + \frac{z^2}{l^2}\ln\frac zl {\tilde g}_{\mu\nu}
+ \ldots\,,
\eeq
which implies\footnote{We use the gauge $\Psi_z = 0$.}
\begin{displaymath}
\Phi = \Phi_0 + \frac{z^2}{l^2}\Phi_2 + \frac{z^4}{l^4}\Phi_4 + \frac{z^4}{l^4}
\ln\frac zl \tilde\Phi + \ldots\,,
\end{displaymath}
\begin{displaymath}
\Psi_{\alpha} = \Psi_{\alpha}^{(0)} + \frac{z^2}{l^2}\Psi_{\alpha}^{(2)} +
\frac{z^4}{l^4}\Psi_{\alpha}^{(4)} + \frac{z^4}{l^4}\ln\frac zl \tilde\Psi_{\alpha}
+ \ldots\,,
\end{displaymath}
\begin{displaymath}
\Omega_{xy} = \Omega_{xy}^{(0)} + \frac{z^2}{l^2}\Omega_{xy}^{(2)} +
\frac{z^4}{l^4}\Omega_{xy}^{(4)} + \frac{z^4}{l^4}\ln\frac zl\tilde\Omega_{xy}
+ \ldots\,,
\end{displaymath}
where all coefficients are functions of $u,x,y$, and we defined
\begin{displaymath}
\tilde\Omega_{xy} = \p_x\tilde\Psi_y - \p_y\tilde\Psi_x\,, \qquad
\Omega_{xy}^{(2n)} = \p_x\Psi_y^{(2n)} - \p_y\Psi_x^{(2n)}\,, \quad
n = 0,1,\ldots
\end{displaymath}
Equ.~(\ref{einsteinui}) yields then
\begin{displaymath}
\tilde\Psi_{\alpha,\alpha} = \Psi^{(2)}_{\alpha,\alpha} = \Psi^{(4)}_{\alpha,\alpha}
= 0\,,
\end{displaymath}
and
\begin{displaymath}
\Omega^{(0)}_{xy,\alpha} = \frac 4{l^2}\epsilon_{\alpha\beta}\Psi^{(2)}_{\beta}\,,
\qquad
\Omega^{(2)}_{xy,\alpha} = -\frac 4{l^2}\epsilon_{\alpha\beta}\tilde\Psi_{\beta}\,.
\end{displaymath}
The Fefferman-Graham expansion of the gauge field $A_u$ is obtained from
(\ref{omegasusy})\footnote{Here we are interested in the supersymmetric case only.},
and reads
\begin{displaymath}
A_u = \frac l{4\sqrt 3}\Omega^{(0)}_{xy} + \frac{z^2}{l^2}A_u^{(2)} -
\frac l{2\sqrt 3}\frac{z^2}{l^2}\ln\frac zl \Omega^{(2)}_{xy} - \frac l{4\sqrt 3}
\frac{z^4}{l^4}\left(\Omega^{(4)}_{xy} - \frac 12\tilde\Omega_{xy}\right) -
\frac l{4\sqrt 3}\frac{z^4}{l^4}\ln\frac zl \tilde\Omega_{xy} + \ldots\,,
\end{displaymath}
where
\begin{displaymath}
A^{(2)}_{u,\alpha} = \frac 1{2\sqrt 3 l}\epsilon_{\alpha\beta}(4\Psi^{(4)}_{\beta}
                     + \tilde\Psi_{\beta})\,.
\end{displaymath}
Finally, the generalized Siklos equation (\ref{siklos}) yields
\begin{displaymath}
\Delta^{(2)}\Phi_0 - \left(\Omega^{(0)}_{xy}\right)^2 - 2\p_u
\Psi^{(0)}_{\alpha,\alpha} - \frac 4{l^2}\Phi_2 = 0\,.
\end{displaymath}
Plugging these results into equations (\ref{hatTuu}), (\ref{hatTux}) and
(\ref{hatTuy}), and subtracting the logarithmically divergent terms that
cannot be removed by adding local counterterms \cite{deHaro:2000xn}, one finally
obtains the holographic stress tensor
\eq
\hat T_{uu} = \frac{N^2}{8\pi^2 l^2}\left[\frac 4{l^2}\Phi_4 + \frac 1{l^2}
              \tilde\Phi - \frac 12 \Delta^{(2)}\Phi_2 + \Omega^{(0)}_{xy}
              \Omega^{(2)}_{xy}\right]\,, \label{hatTuuFG}
\eeq
\eq
\hat T_{u\alpha} = \frac{N^2}{16\pi^2 l^2}\epsilon_{\alpha\beta}\p_{\beta}
                   \left[\Omega^{(2)}_{xy} - \frac{4\sqrt 3}l A_u^{(2)}\right]\,.
\eeq
Using the CFT metric
\eq
ds^2 = -2\,du\,dv + \Phi_0\,du^2 + 2\left(\Psi^{(0)}_x dx + \Psi^{(0)}_y dy
\right)du + dx^2 + dy^2\,, \label{cftmetric}
\eeq
one easily checks that this energy-momentum tensor is covariantly conserved,
$\nabla_a \hat T^a_{\;\;\,b} = 0$. Note that (\ref{cftmetric}) describes itself a
gyraton (in flat space), unless the functions $\Phi_0$ and $\Psi^{(0)}_{\alpha}$
vanish. (For $\Psi^{(0)}_{\alpha} = 0$ and $\Phi_0 \neq 0$ the CFT would live on
a pp-wave background).

For the Kaigorodov spacetime, which is given by $\Psi_{\alpha} = 0$ and
$\Phi = z^4\Phi_4/l^4$ with constant $\Phi_4$, (\ref{hatTuuFG}) reduces to the
result found in \cite{Brecher:2000pa}, namely that the dual CFT has constant null
momentum density. We see that in our case one has in addition a nonvanishing
angular momentum density. It would be interesting to explore the AdS/CFT
interpretation of our gyratons in more detail.

\section{Final remarks}

In this paper we obtained gyraton solutions of minimal gauged and ungauged
supergravity in five dimensions and analyzed under which conditions
they preserve some supersymmetry. It was shown that the gyratons on AdS
enjoy a Siklos-Virasoro reparametrization invariance, and that the holographic
stress tensor associated to these solutions transforms without anomaly
under these transformations, contrary to claims that appeared previously in
the literature \cite{Banados:1999tw}.

We furthermore obtained supersymmetric gyratons in both gauged and ungauged
$N=2$ five-dimensional supergravity coupled to an arbitrary number of
vector multiplets. It would be interesting to study the asymptotically AdS
gyratons from the point of view of the AdS/CFT correspondence. In this context
it has been conjectured that gravity on the Kaigorodov spacetime, which
represents a special case of a pp-wave in AdS, is dual to a conformal field
theory in the infinite momentum frame with constant momentum
density \cite{Cvetic:1998jf}.

Another interesting point would be to look for gyraton solutions in
ten- and eleven-dimensional supergravity and to see how much supersymmetry
they preserve. Since string theory on the maximally supersymmetric
IIB plane wave background is exactly quantizable \cite{Metsaev:2002re},
one could ask the question if this is still the case for a gyraton background,
which represents a generalization of the pp-wave. Work in this direction is
in progress \cite{ckz}.

\acknowledgments

This work was partially supported by INFN, MURST and
by the European Commission program
MRTN-CT-2004-005104.

\end{document}